\documentclass[iop]{emulateapj}
\usepackage{amsmath}
\usepackage{amssymb}
\usepackage{graphicx}
\usepackage{epstopdf}
\usepackage{natbib}
\usepackage{color}

\begin{document}

\newcommand{\etal}{et~al.}
\newcommand{\PaperOne}{Paper~1}

\title{Size dependence of the radio luminosity -- mechanical power correlation in radio galaxies}
\shorttitle{Radio luminosity -- mechanical power correlation}
\shortauthors{Shabala \& Godfrey}

\author{S. S. Shabala}
\affil{School of Mathematics and Physics, Private Bag 37, University of Tasmania, Hobart, TAS 7001, Australia}
\email{Stanislav.Shabala@utas.edu.au}
\author{L. E. H. Godfrey}
\affil{International Centre for Radio Astronomy Research, Curtin University, GPO Box U1987, Perth, Western Australia 6845, Australia}

\begin{abstract}

We examine the relationship between source radio luminosity and kinetic power in Active Galactic Nucleus (AGN) jets. We show that neglecting various loss processes can introduce a systematic bias in the jet powers inferred from radio luminosities for a sample of radio galaxies. This bias can be corrected for by considering source size as well as radio luminosity; effectively the source size acts as a proxy for source age. Based on a sample of FR-II radio sources with jet powers derived from the measured hotspot parameters, we empirically determine a new expression for jet power that accounts for the source size, $\frac{Q_{\rm jet}}{10^{36} {\rm \,W}} = 1.5^{+1.8}_{-0.8} \left( \frac{L_{\rm 151}}{10^{27}{\rm \,W\,Hz^{-1}}} \right)^{0.8} \left( 1+z \right)^{1.0} \left( \frac{D}{\rm kpc} \right)^{0.58 \pm 0.17}$, where $D$ is source size and $L_{151}$ the 151 MHz radio luminosity. By comparing a flux-limited and volume-limited sample, we show that any derived radio luminosity -- jet power relation depends sensitively on sample properties, in particular the source size distribution and the size-luminosity correlation inherent in the sample. Such bias will affect the accuracy of the kinetic luminosity function derived from lobe radio luminosities and should be treated with caution.

\end{abstract}


\section{Introduction}

There has been much recent interest in accurate measurement of the kinetic power of Active Galactic Nucleus (AGN) jets. Such measurements facilitate studies of AGN physics (for example jet composition, and generation and collimation mechanisms); as well as helping to quantify the effects of AGN feedback on galaxies and their building blocks in a cosmological framework.

AGN jet powers in individual objects have been estimated using cavities in the hot X-ray gas (Rafferty et al. 2006; Birzan et al. 2008; Cavagnolo et al. 2010); dynamical models of radio sources (Machalski et al. 2004; Shabala et al. 2008); and radio lobe expansion speeds (Daly et al. 2012). Recently, in a companion paper (Godfrey \& Shabala 2012, hereafter Paper~1) we presented a new method based on measured properties of jet termination shocks known as hotspots. This method agrees well with other techniques.

It is highly desirable to develop a very simple method of measuring AGN jet power, ideally one that relies on a minimal number of easily measurable observables. The AGN radio luminosity function (RLF) is the standard measurement used to study cosmological evolution of the AGN population. Various authors (e.g. Willott et al. 1999; Heinz \& Sunyaev 2003; Merloni \& Heinz 2007; Cavagnolo et al. 2010) have attempted to convert radio luminosity, $L_{\rm radio}$, to AGN mechanical power, $Q_{\rm jet}$. Such a conversion is not easy, since the observed radio luminosity depends on the details of the electron acceleration and loss processes, as well as interactions between the radio source and its environment. However, the stakes are high, since a correct $L_{\rm radio} - Q_{\rm jet}$ relationship allows the whole radio luminosity function to be converted to a kinetic jet power function (e.g. Merloni \& Heinz 2007; Mart\'inez-Sansigre \& Rawlings 2011; Kaiser \& Best 2007; Cattaneo \& Best 2009), and therefore AGN heating rate. Cattaneo \& Best (2009) show that the uncertainty in the assumed $L_{\rm radio} - Q_{\rm jet}$ relation dominates the uncertainty in the kinetic jet power function. It is therefore of fundamental importance to determine the $L_{\rm radio} - Q_{\rm jet}$ relation, or relations. The ultimate goal is to self-consistently reconcile in the framework of galaxy formation models the observed AGN properties (i.e. the RLF) with galaxy and interstellar/intracluster medium properties (mass functions, colours).

One of the most commonly used relations in converting radio luminosity to jet power is that presented by Willott et al. (1999). Based on synchrotron minimum energy calculations in combination with a self-similar model of radio galaxy evolution (Falle 1991; Kaiser \& Alexander 1997), these authors predicted $Q_{\rm jet} \propto L_{\rm radio}^{6/7}$. This proportionality was supported by an observed correlation between [O\,II] narrow-line luminosity, assumed to be a proxy for the continuum ionising luminosity and therefore related to jet power\footnote{As we point out in Paper 1, the slope of the $L_{\rm OII} - L_{151}$ correlation is highly uncertain and sample dependent (e.g. Hardcastle et al. 2009, Fernandes et al. 2011). Furthermore, how well the narrow-line luminosity traces jet power is highly debatable (compare e.g. Hardcastle et al. (2009) and Tadhunter et al. (1998) with Rawlings \& Saunders (1991) and Shabala et al. (2012)).}, and 151 MHz lobe radio luminosity. Unlike many other correlations (e.g. B\^irzan \etal\/ 2008; Daly \etal\/ 2012) this result is not purely empirical: the correlation can be understood in terms of a dynamical radio source model under some quite reasonable assumptions.

In this work, we use the hotspot method described in Paper~1 to re-examine the $Q_{\rm jet} - L_{\rm radio}$ relation of Willott et al. (1999). We show that the conversion of radio luminosity to jet power is sensitive to the size of the AGN radio lobes, largely due to inverse Compton losses. We note that the analysis of Willott et al. (1999) neglected this important radiative loss mechanism, which becomes increasingly relevant for old ($>100$~Myrs) AGN, resulting in a systematic shift in jet power estimates. As a result of these losses, the slope and normalisation of the $Q_{\rm jet} - L_{\rm radio}$ relation are dependent on the distribution of source sizes within the sample being considered.

Throughout the paper, we adopt the concordance cosmology of $\Omega_{\rm M}=0.27$, $\Omega_\Lambda=0.73$, $h=0.71$.

\section{Jet power from source dynamics}

\subsection{Radio source dynamics}
\label{sec:dynModel}

The class of dynamical models of double-lobed Fanaroff-Riley Type II radio sources, pioneered by Scheuer (1974), and subsequently developed by Begelman \& Cioffi (1989), Falle (1991) and Kaiser \& Alexander (1997) among others, describe the evolution of an overpressured cocoon of radio plasma. Jets of synchrotron--emitting plasma emanating from the black hole accretion disk smash into the ambient gas forming termination shocks; observationally these are manifested as bright regions of radio emission, known as hotspots. The overpressured hotspot plasma then flows backwards, inflating a cocoon of radio--emitting plasma. Kaiser \& Alexander (1997) argued that the pressure from the cocoon may be responsible for keeping AGN jets collimated over long distances. These authors showed that, in this case, the radio source will necessarily expand self-similarly, its size growing as 
 
\begin{equation} \label{eqn:Ddynamical}
D = 2 c_1 R_0 \left( \frac{Q_{\rm jet}}{\rho_0 R_0^\beta} \right)^{\frac{1}{5-\beta}} t^{\frac{3}{5-\beta}}
\end{equation}
where $c_1$ is a constant of order unity (see Appendix), $Q_{\rm jet}$ is the jet kinetic power; and the source is assumed to expand into an atmosphere with density profile $\rho(r)=\rho_0 \left( \frac{r}{R_0} \right)^{-\beta}$.

The observed radio luminosity is produced by synchrotron emitting electrons and/or positrons in the cocoon. Thus, the luminosity depends on the cocoon magnetic field energy density $u_B$, the distribution $n(\gamma)$ of the Lorentz factors of the emitting electrons, and cocoon volume $V$. In the absence of loss processes, it can be shown (e.g. Longair 1994) that, for an initial electron energy distribution $n(\gamma_i) =  k_e  \gamma_i^{-p}$ injected at the hotspot between a minimum and maximum Lorentz factor $\gamma_1$ and $\gamma_2$, under the assumption of equipartition between particle and magnetic field energy density, the cocoon radio luminosity at frequency $\nu$ is 

\begin{equation} \label{eqn:Lradio_u_B}
L_{\rm \nu} = A_1 u_B^{\frac{5+p}{4}} V \nu^{\frac{1-p}{2}}  (1+z)^{\frac{(3-p)}{2}}
\end{equation}
where $A_1$ is a numerical constant given by Equation \ref{eqn:A_1} in the Appendix. For our fiducial values of $p=2.5$, $\gamma_1 = 10^3$ and $\gamma_2 = 10^5$, we have $A_1 = 3.5 \times 10^{-8}$. 

For self-similar expansion, the cocoon volume $V = \pi R_{\rm T}^2 D^3$ where $R_{\rm T}$ is the axial ratio of the source. Again, with the assumption of equipartition between particle and magnetic field energies as the radio source expands, the magnetic field energy density is proportional to cocoon pressure, $u_B = p_{\rm c} \frac{r}{(r+1)(\Gamma_l - 1)}$ where $r=\frac{p+1}{4}$ and $\Gamma_l$ is the adiabatic index of the lobes \citep{KaiserBest07, KaiserBest07_erratum}. \citet{KA97} give an expression for the evolution of cocoon pressure $p_{\rm c}$ with time (their Equation~20, see also \citet{KaiserBest07} Equation A4), which yields 

\begin{equation} \label{eqn:u_B}
 u_B = \frac{r f_p}{(r+1)(\Gamma_l -1)} \left( \rho_0 R_0^\beta Q_{\rm jet}^2 \right)^{1/3} D^{-(4+\beta)/3} 
\end{equation}
where $f_p$ is a constant given in Equation A5 of \citet[][]{KaiserBest07} (see also the Appendix). The radio luminosity is then 
\begin{eqnarray}
L_{\rm \nu} &=& A_1 \pi R_{\rm T}^2 \left(  \frac{r f_p}{(r+1)(\Gamma_l -1)}  \right)^{\frac{(5+p)}{4}} \nu^{(1-p)/2} \nonumber \\ 
&\times& \left( 1+z \right)^{\frac{3-p}{2}} \left( \rho_0 R_0^\beta Q_{\rm jet}^2 \right)^{\frac{5+p}{12}} D^{3-\left( \frac{4+\beta}{3} \right) \left( \frac{5+p}{4} \right)} 
\label{eqn:LradioLosslessGeneral}
\end{eqnarray}

Note that the synchrotron spectral index is just $\alpha=(1-p)/2$. A reasonable hotspot spectral index between $\alpha=-0.5$ and $-0.75$ gives $p=2.0-2.5$. Because of radiative cooling discussed below, the observed cocoon spectral index is typically steeper. Taking $p=2$ and assuming a reasonable value of $\beta=1.5$ (e.g. Vikhlinin et al. 2006)\footnote{A less conservative range of $1 < \beta < 2.5$, $2<p<2.7$ yields a range of [-1.2, 0.1] for the exponent of $D$ in Equation~\ref{eqn:LradioLosslessGeneral}.} yields 
\begin{equation}
L_{\rm \nu} \propto \left( \rho_0 R_0^\beta Q_{\rm jet}^2 \right)^{7/6} D^{-5/24}
\label{eqn:LradioLosslessWillott}
\end{equation}
For this set of assumptions, at a fixed radio luminosity, the relationship between jet power and size is then $Q_{\rm jet} \propto D^{5/28}$. As Willott et al. (1999) point out, for realistic values of $D$ (typically $1-1000$~kpc) this is a very weak dependence. Hence, one expects $Q_{\rm jet} \propto L_{\rm radio}^{6/7}$ for powerful radio galaxies. We note that, as Equation~\ref{eqn:LradioLosslessGeneral} clearly shows, the $Q_{\rm jet} - L_{\rm radio}$ relation depends on the source spectral index.

\subsection{Size dependence of luminosity}
\label{sec:dynModelLosses}

The above analysis is only correct if losses in the synchrotron--emitting population are negligible. Three loss mechanisms conspire to modify the electron energy spectrum: adiabatic expansion of the volume element in which the electrons were last accelerated, synchrotron radiation losses, and upscattering of Cosmic Microwave Background photons via the inverse Compton process. As a result of these various loss mechanisms, the Lorentz factor evolves as ({\em cf} Kaiser et al. 1997 Equation 4)

\begin{equation}
	\frac{d \gamma}{dt} = -\frac{a_2}{3} \frac{\gamma}{t} - \frac{4}{3}\frac{\sigma_T}{m_e c} \gamma^2 (u_B + u_C)
\label{eqn:dgamma_dt}
\end{equation}
where $a_2=\frac{5}{3} \left( \frac{4+\beta}{5-\beta} \right)$, $u_C = 4 \times 10^{-14} (1+z)^4$~J\,m$^{-3}$ is the energy density of the CMB photons, and the other symbols have their usual meanings.

To a good approximation, the observed radio luminosity comes from electrons emitting synchrotron radiation at  $\nu=\gamma^2 \nu_L$, where $\nu_L$ is the Larmor frequency. The evolution of $\gamma(t)$ for a population of electrons with initial Lorentz factor $\gamma_i$ at acceleration time $t_i$ is then given by Equation~\ref{eqn:dgamma_dt}. As shown by Kaiser et al. (1997), integrating over all injection times then yields the radio luminosity as a function of source age. An important point is that the radio luminosity will evolve with age, as electrons radiate away their energy. This is not taken into account in the simple analysis of Section~\ref{sec:dynModel}.

\section{Results}

\subsection{Comparing jet power estimates}
\label{sec:hs_vs_dyn}

\begin{figure}
\begin{center}
\includegraphics[width=0.33\textwidth,angle=270]{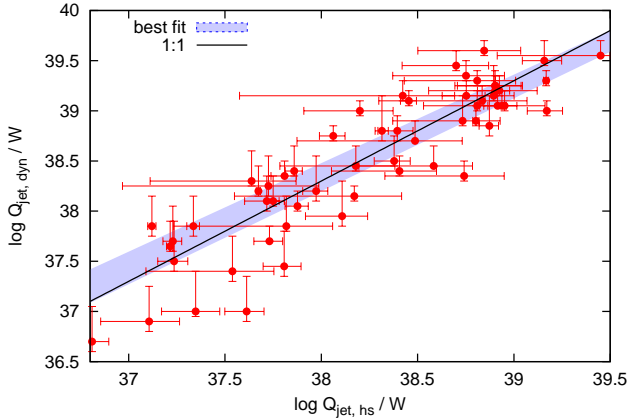}
\end{center}
\caption{Comparison of jet power estimates for the hotspot sample. Hotspot jet powers are determined for $g=2$ (see Paper 1), and are independent of radio source environment and orientation; while dynamical models assume the same environment, shape and orientation for all sources. These assumptions contribute to the scatter. Model parameters are as follows: cocoon axial ratio $R_{\rm T}=2.3$,  the median value in our sample; minimum Lorentz factor $\gamma_{\rm min}=1000$ \citep{GodfreyEA09}; and particle injection index $p=2.5$. The density profile is as in Willott et al. (1999), with $\beta=1.5$ and particle density at 100~kpc $n_{\rm 100}=3000$~m$^{-3}$. Shaded region shows the best fit between these estimates. The solid line indicates the expected 1:1 correspondence.}
\label{fig:Qjet_hs_vs_dyn}
\end{figure}

Figure~\ref{fig:Qjet_hs_vs_dyn} shows a comparison of hotspot and dynamical jet powers for the sample described in {\PaperOne}. These objects were selected from a complete $z<1$ sample of FR~II radio sources \citep{MullinEA08} on the basis of having at least one (and often two) well-resolved terminal hotspots, allowing their jet powers to be determined. For details of the sample selection procedure and jet power calculations, we refer the reader to {\PaperOne}. 

Dynamical jet powers were determined from source size and luminosity using a grid of dynamical models. These are generated as described in Shabala et al. (2008). Briefly, for each jet power, a source size and luminosity are calculated for each point in time given an environment into which the radio source is expanding. Once a model grid is generated, the observed sizes and luminosities are used to infer dynamical jet powers. Our modeling assumes all sources expand into the same environment, and have the same shape (i.e. axial ratio $R_{\rm T}$ relating cocoon length to its width). Formal uncertainties in dynamical jet power estimates are largely due to uncertainty in the source orientation to the observer's line of sight. Error bars on hotspot jet power estimates correspond to the difference in derived jet powers in the two hotspots (see {\PaperOne} for details). While axial ratio values are available for individual sources, we chose not to use these in order to assess the sensitivity of dynamical jet power estimates to this parameter. Together with the ``same environment'' assumption, this yields some scatter in the $Q_{\rm jet,dyn} - Q_{\rm jet,hs}$ relation. However, there is a strong correlation consistent with a linear relationship.

\subsection{Size--luminosity tracks}

In the following results we further restrict our sample by requiring that jet powers from each hotspot pair agree within a factor of two. This ensures that the measured hotspot jet powers are closely related to total radio source size and luminosity, rather than being dependent on small-scale ISM inhomogeneities in or near the hotspot region. Our sample comprises of 44 radio sources.

\begin{figure}
\begin{center}
\includegraphics[width=0.33\textwidth,angle=270]{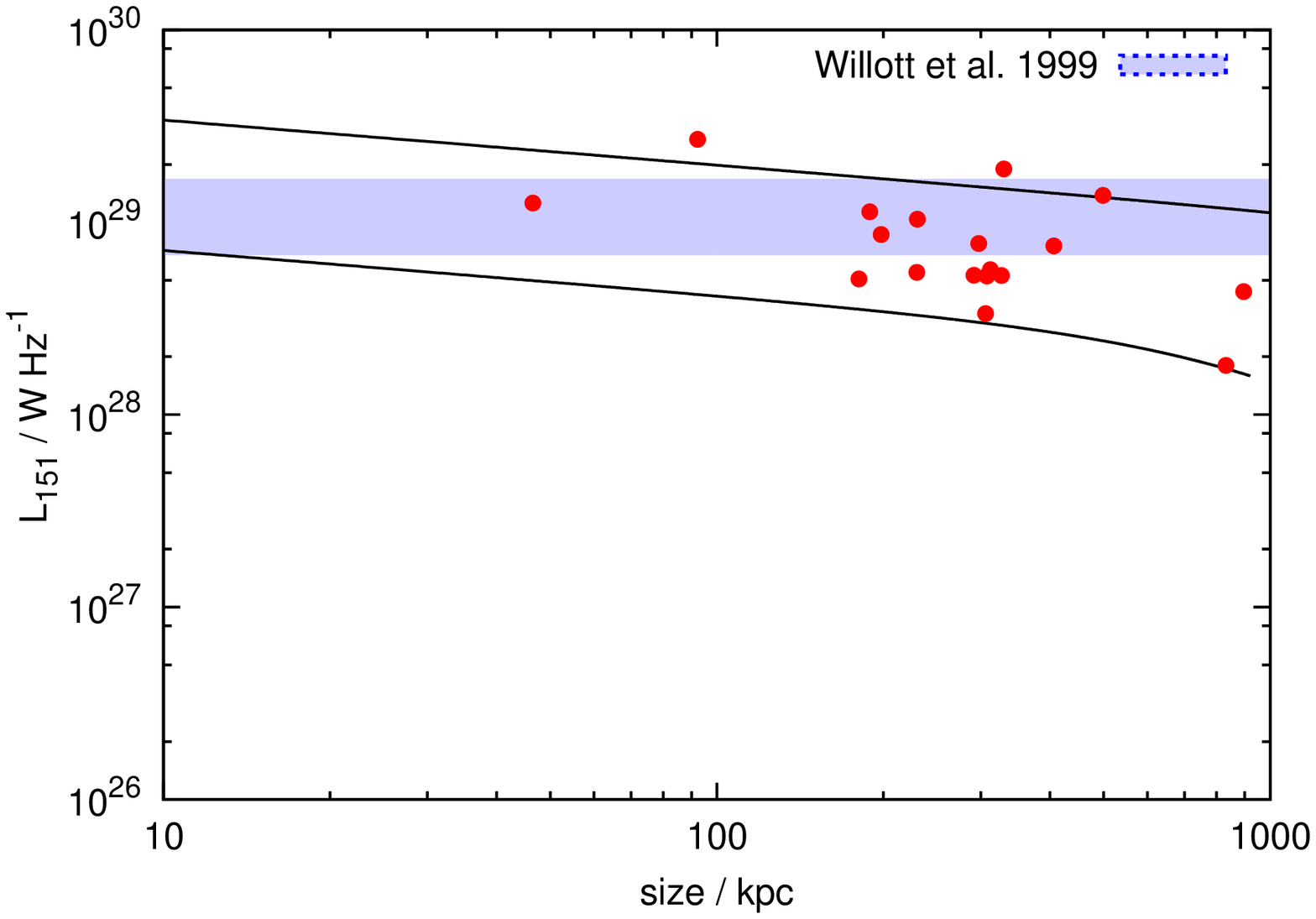}
\includegraphics[width=0.33\textwidth,angle=270]{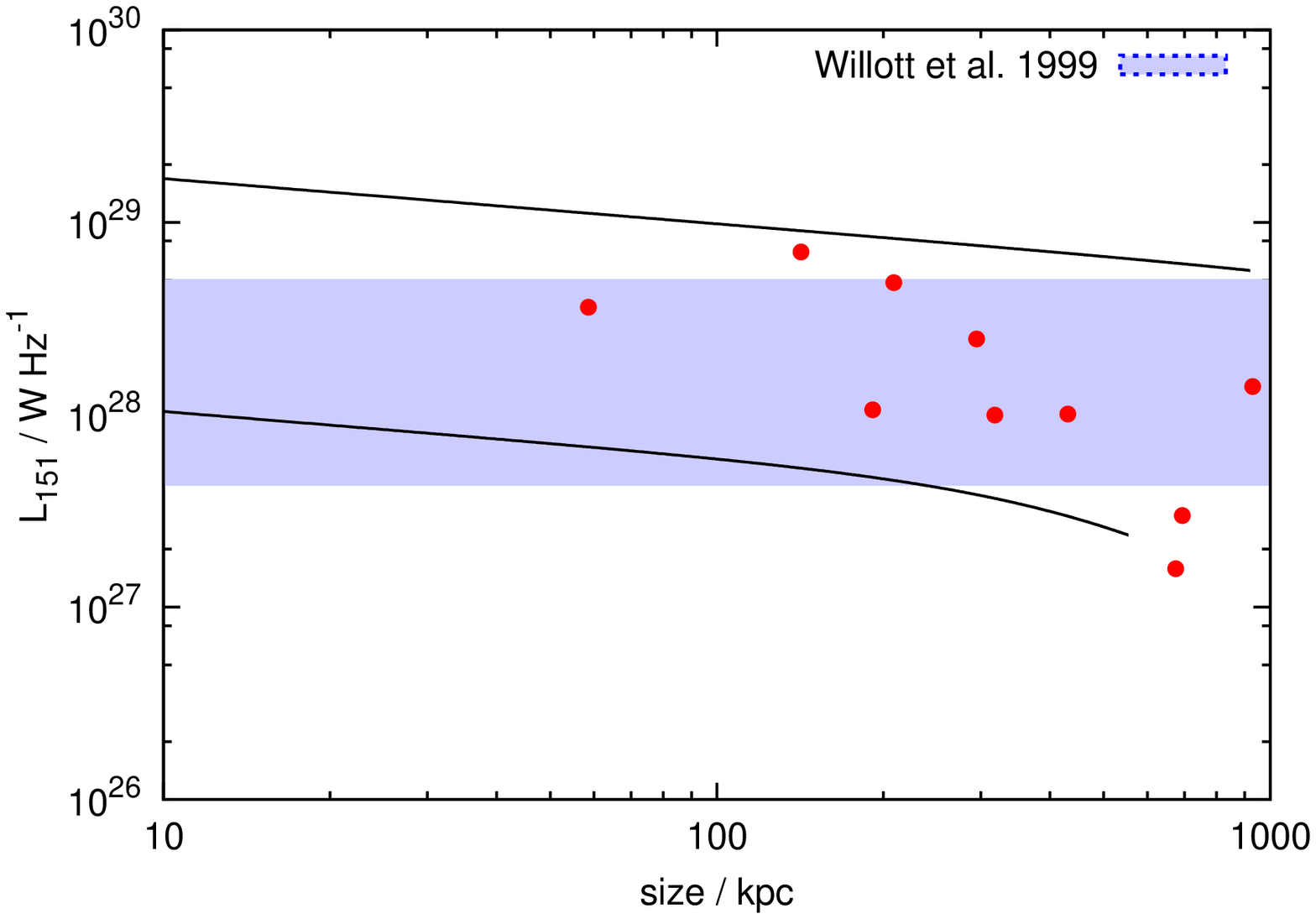}
\includegraphics[width=0.33\textwidth,angle=270]{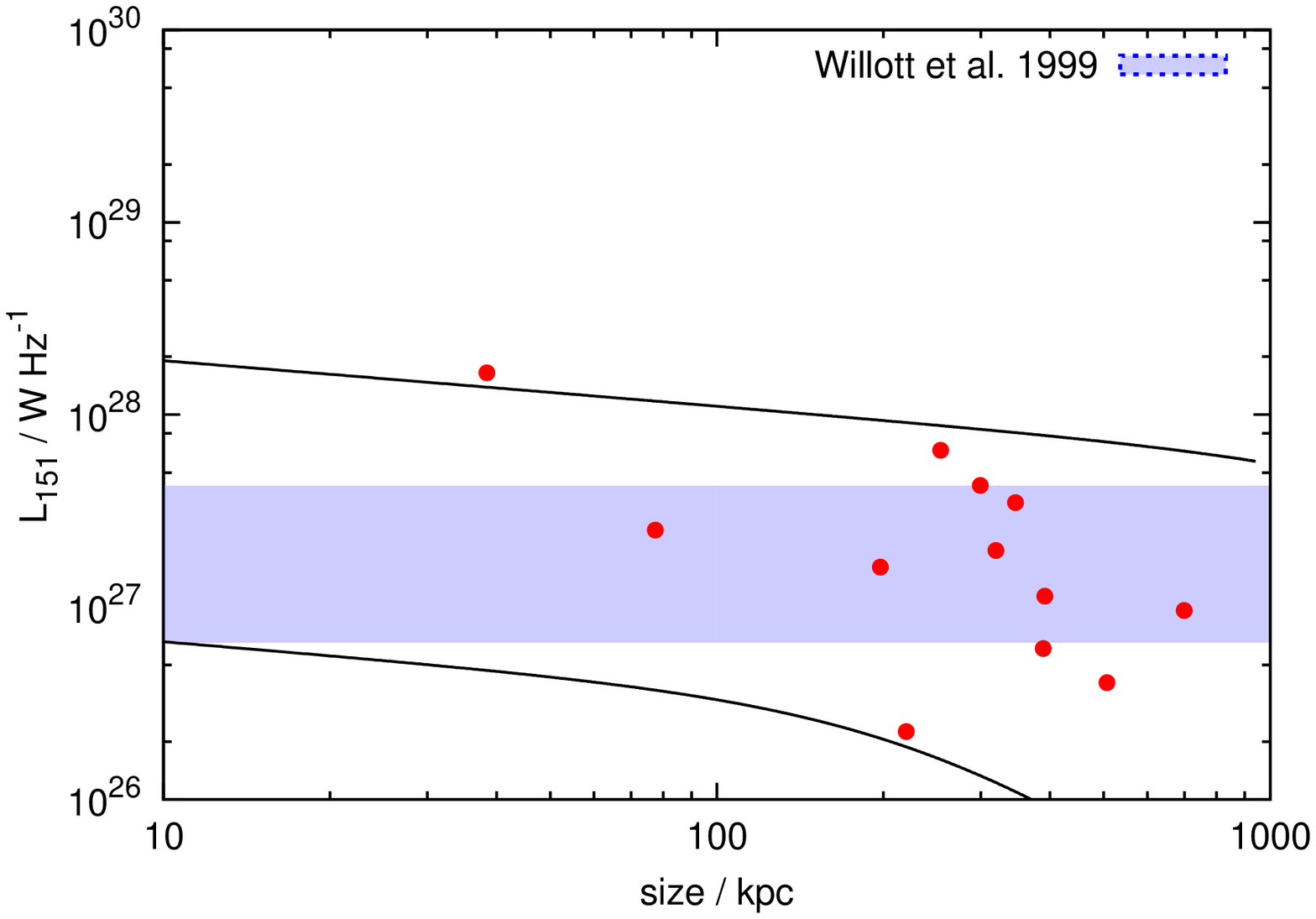}
\end{center}
\caption{Size--luminosity distribution of FR-II radio sources with well-defined hotspots. Jet powers are in the range $6 \times 10^{38}$--$3 \times 10^{39}$ (left), $6 \times 10^{37}$--$6 \times 10^{38}$ (middle), and $6 \times 10^{36}$--$6 \times 10^{37}$~W (right panel). In each case, model tracks (solid lines) show the predicted evolution of highest and lowest jet power sources. For all plots, the parameters are as in Figure~\ref{fig:Qjet_hs_vs_dyn}. Only the jet power is varied between panels. Predictions of the Willott et al. (1999) relation are shown with a shaded region.}
\label{fig:PDtracksHotspots}
\end{figure}

Figure~\ref{fig:PDtracksHotspots} shows the size--radio luminosity tracks for this sample. We use hotspot jet powers to split our sample into high jet power (left panel), medium (middle) and low jet power (right) bins. Solid lines show the limiting expected size--luminosity tracks. For each panel, the lower track is obtained by taking the lowest jet power source in that bin, and evolving it at the highest redshift in the sample\footnote{The redshift dependence comes from the conversion between the observed and emitted frequency, $\nu_{\rm obs}= \nu_{\rm em} / (1+z)$. As radio lobes are steep-spectrum sources, the radio luminosity at a given observing frequency decreases with redshift. There is also a redshift dependence due to the changing CMB energy density ($u_C \propto (1+z)^4$) in Equation~\ref{eqn:dgamma_dt}.} ($z=1$). Conversely, each upper track is obtained by evolving the highest jet power source at the lowest redshift ($z=0$).

Two things are apparent from Figure~\ref{fig:PDtracksHotspots}. First, at a fixed jet power there is a clear dependence of radio luminosity on source size. The shaded regions show radio luminosities expected from the size-independent relation of Willott et al. (1999; their Equation 12). Second, dynamical models do a very good job of reproducing this evolution of radio luminosity with size. It is worth emphasising that the only parameter allowed to change between the three panels in Figure~\ref{fig:PDtracksHotspots} is jet power, which is determined from a completely orthogonal set of (hotspot) observations. All other parameters, such as those describing the density profile of the atmosphere into which the radio source is expanding, radio source shape, and initial particle energy distributions, are held fixed. The decrease in radio luminosity with size is easily understood in terms of loss processes described by Equation~\ref{eqn:dgamma_dt}.

\subsection{Mechanical power -- radio luminosity -- size relation}
\label{sec:fit_LQD}

We can combine the three panels of Figure~\ref{fig:PDtracksHotspots} to elucidate the dependence of jet power on source size. Combining expressions for source size (Equation~\ref{eqn:Ddynamical}) and radio luminosity (Equation~\ref{eqn:LradioLosslessGeneral}) yields
\begin{equation}
\left[ \frac{4}{5+p} \right] \log L_{\rm radio} - \frac{2}{3} \log Q_{\rm jet} = A_2 + \left[ \frac{5-\beta}{3} - \frac{3 (1+p)}{5+p} \right] \log D
\label{eqn:LQD_scaling}
\end{equation}
where $A_2$ is a normalisation constant given in the Appendix. The left-hand side of this equation represents a radio luminosity scaled by jet power in accordance with the dynamical model of \citet{KA97}. For our adopted parameters, Equation~\ref{eqn:LQD_scaling} reduces to $0.533 \log L_{\rm radio} - \frac{2}{3} \log Q_{\rm jet} = A_2 - 0.233 \log D$. Because our sample spans a large redshift range ($z=0-1$), the radio luminosity must be adjusted accordingly. This is difficult to do analytically, but can be readily done with dynamical models. For our parameters, this yields a correction of the form\footnote{This is the correction required to match the $z=0$ and $z=1$ size-luminosity tracks for the range of jet powers in our sample.} $L_{\rm radio}(z)  = L_{{\rm radio, } z=0} (1+z)^{1.27}$. 

\begin{figure}
\begin{center}
\includegraphics[width=0.33\textwidth,angle=270]{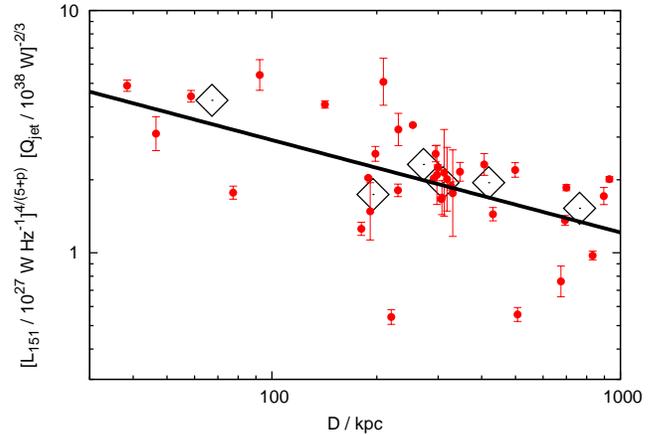}
\end{center}
\caption{Size--luminosity tracks of Figure~\ref{fig:PDtracksHotspots} adjusted for jet power and redshift (see text). Only sources with $Q_{\rm jet} > 10^{44}$\,ergs s$^{-1}$ are considered in this analysis. The \citet{WillottEA99} relation predicts a horizontal line, ruled out by our Monte Carlo simulations at the 1 percent level. Medians are shown by open diamonds, and the best-fit relation of Equation~\ref{eqn:LQD_bestFit} by the solid line.}
\label{fig:PDtracks_scaled}
\end{figure}

Figure~\ref{fig:PDtracks_scaled} shows the dependence of the radio luminosity -- jet power combination of Equation~\ref{eqn:LQD_scaling} on source size. We note that if loss processes associated with the radiating particles were not important, as is the case in the relation given by \citet{WillottEA99}, we would expect all points to lie on a horizontal line. In other words, there would be no size dependence of the ordinate. We performed 10,000 Monte-Carlo bootstrap resampling realizations of our data, adopting an average uncertainty of $\sigma_{\rm log Q_{\rm jet}}=0.2$~dex in individual jet power measurements. This value comes from the observed distribution of north-to-south hotspot jet power ratios, plotted as error bars in Figure~\ref{fig:PDtracks_scaled}. Our Monte-Carlo simulations yielded a slope of $-0.38 \pm 0.11$ and intercept of $1.43 \pm 0.28$. The derived slope is significantly different from zero at the 1 percent level, ruling out the null hypothesis of no size dependence. As we explain in the following section, this result can be readily understood by considering the energy loss processes of the radiating particles.

We use our data to provide a new estimator of jet power from radio luminosity and source size,

\begin{eqnarray}
\frac{Q_{\rm jet}}{10^{36} {\rm \,W}} &=& 1.5^{+1.8}_{-0.8} \left( \frac{L_{\rm 151}}{10^{27}{\rm \,W\,Hz^{-1}}} \right)^{0.8} \nonumber \\
&& \times \left( 1+z \right)^{1.0} \left( \frac{D}{\rm kpc} \right)^{0.58 \pm 0.17}
\label{eqn:LQD_bestFit}
\end{eqnarray}
where we have assumed a hotspot jet power normalisation constant $g=2$, consistent with our findings in Paper~1. 

As we discuss below, the size term in Equation~\ref{eqn:LQD_bestFit} plays an important role in the conversion of radio luminosity to jet power. In particular, flux and volume-limited samples can yield very different scaling relations, due to the different correlations between source size and radio luminosity inherent in each sample.
 
\section{Discussion}

\subsection{Spectral ageing}
\label{sec:specAgeing}

The size dependence of the observed $Q_{\rm jet} - L_{\rm radio}$ relation can be understood by considering the predicted relation in the lossless case (Equation~\ref{eqn:LradioLosslessWillott}). Re-writing this in terms of the spectral index $\alpha=(1-p)/2$ gives $L_{\rm radio} \propto Q_{\rm jet}^{(3-\alpha)/3} D^{\alpha \left( \frac{4+\beta}{6} \right) + \left( 1 - \frac{\beta}{2} \right)}$. This is similar to the expression of O'Sullivan et al. (2011), but also includes a size term. Rearranging,

\begin{eqnarray}
%Q_{\rm jet} &=& \left(  A_1 \pi R_{\rm T}^2 \right)^{-\frac{3}{3-\alpha}}   \left( \frac{r f_p}{(r+1)(\Gamma_l - 1)}  \right)^{-3}   \left(  \rho_0 R_0^\beta \right)^{-\frac{1}{2}}  \nu^{\frac{3 \alpha}{3 - \alpha}} \nonumber \\
%&\times& L_{\rm radio}^{3/(3-\alpha)} D^{\left( 2+\frac{\beta}{2} \right) - \left( \frac{9}{3-\alpha} \right) }
Q_{\rm jet} &=& A_3 L_{\rm radio}^{3/(3-\alpha)} D^{\left( 2+\frac{\beta}{2} \right) - \left( \frac{9}{3-\alpha} \right) }
\label{eqn:QjetLradio_spIndex}
\end{eqnarray}
where $A_3$ is a function that absorbs the other model parameters and numerical constants and is given in Equation \ref{eqn:A_3} of the Appendix. As the radio source ages, loss processes discussed in Section~\ref{sec:dynModelLosses} will cause the observed lobe spectral index $\alpha_{\rm lobe}$ to steepen to a value less than the injected spectral index $\alpha_{\rm inj}$ (note that ``normal'' spectral indices are negative in our notation). This is a well-known observational result: for example, Blundell et al. (1999) find strong evidence for large sources having steeper spectral indices (see their Fig. 8), with sources larger than 100 kpc often having $\alpha_{\rm lobe} \leq -1.0$, compared to $\alpha_{\rm lobe} \sim -0.5$ for sub-100 kpc sources. This will affect the $Q_{\rm jet} - L_{\rm radio}$ relation in two ways. First, the steeper spectral index will affect the luminosity term, changing the slope from $3/(3-\alpha)=0.87$ (for $\alpha=-0.5$) to $3/(3-\alpha)=0.75$ (for $\alpha=-1$). Crucially, there will also be an additional size dependence. As argued by Willott et al. (1999), and in Section~\ref{sec:dynModel}, this dependence is very weak for sources that have not suffered significant losses ($\alpha_{\rm lobe} \approx \alpha_{\rm inj} =-0.5$), with the exponent of $D$ in Equation~\ref{eqn:QjetLradio_spIndex} $\sim 0.2$. Steepening of the spectral index to $\alpha = -1$ makes the exponent of D equal to $0.5$, consistent with the dependence found in Equation~\ref{eqn:LQD_bestFit}. Importantly, for such sources an order of magnitude change in size (e.g. from 30 to 300 kpc) changes the normalisation of the $Q_{\rm jet} - L_{\rm radio}$ relation by a factor of 3, with larger sources having higher jet powers at a given lobe radio luminosity.

\subsection{Effects of sample selection on the mechanical power -- radio luminosity relation}
\label{sec:Qjet_Lradio}

\begin{figure}
\begin{center}
\includegraphics[width=0.33\textwidth,angle=270]{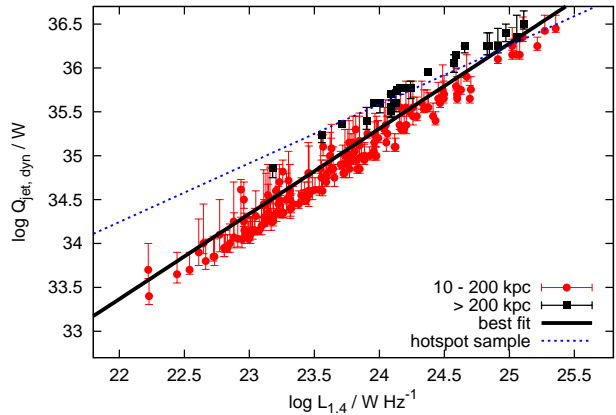}
\end{center}
\caption{Dynamical jet power -- radio luminosity relation for the volume-limited sample of Shabala \etal\/ (2008). Despite significant scatter induced by environment and orientation effects, at a fixed radio luminosity the largest sources have systematically higher jet powers than their more compact counterparts. This systematic shift is primarily due to inverse Compton ageing of the electron population. The best fit $Q_{\rm jet} - L_{1.4}$ relation for all points in this sample is shown by the solid black line. The dashed blue line is the relation derived in Paper~1 for the hotspot sample, consisting predominantly of large ($>100$~kpc) sources. The slope of the best-fit relation for the volume-limited sample is $1.02 \pm 0.02$, compared with $0.67 \pm 0.05$ for the hotspot sample. This difference comes about because of the different size distributions of the sources making up the two samples (Figure~\ref{fig:Qjet_Lradio_sampleSelection}).}
\label{fig:Qjet_vs_Lradio_saar08}
\end{figure}

We illustrate the importance of the size dependence of jet power by considering another sample of radio AGN, a low-redshift ($0.03<z<0.1$) volume-limited sample of Shabala \etal\/ (2008). Figure~\ref{fig:Qjet_vs_Lradio_saar08} shows the relationship between the dynamical jet power and 1.4~GHz cocoon radio luminosity. The sample is split in source size. The largest sources clearly lie above the best-fit relation, shown by the solid line. This is as expected from source dynamics: the oldest (and therefore largest) sources will suffer significant losses, and have a lower radio luminosity for a given jet power, than their more compact counterparts. 

\begin{figure}
\begin{center}
\includegraphics[width=0.33\textwidth,angle=270]{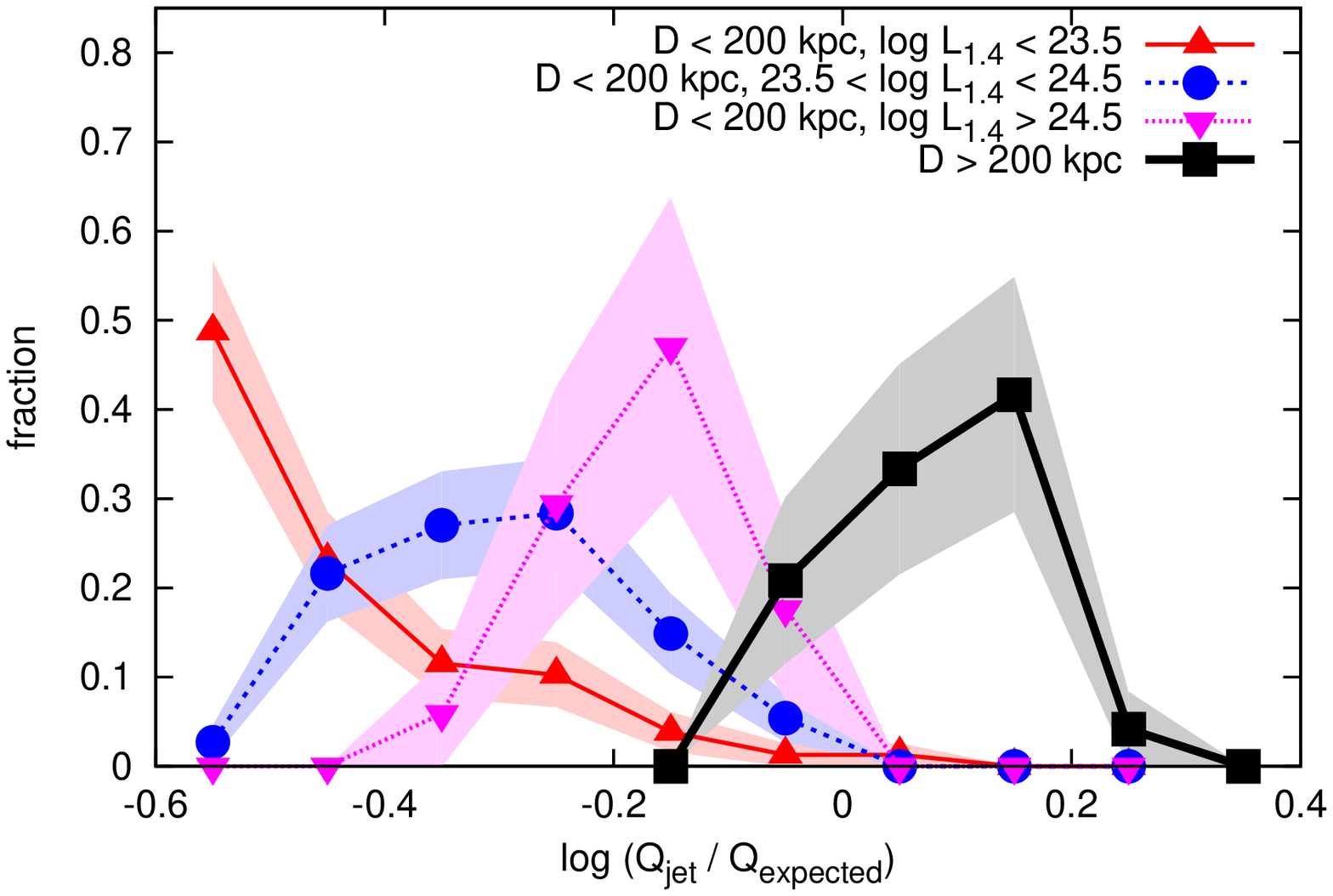}
\includegraphics[width=0.33\textwidth,angle=270]{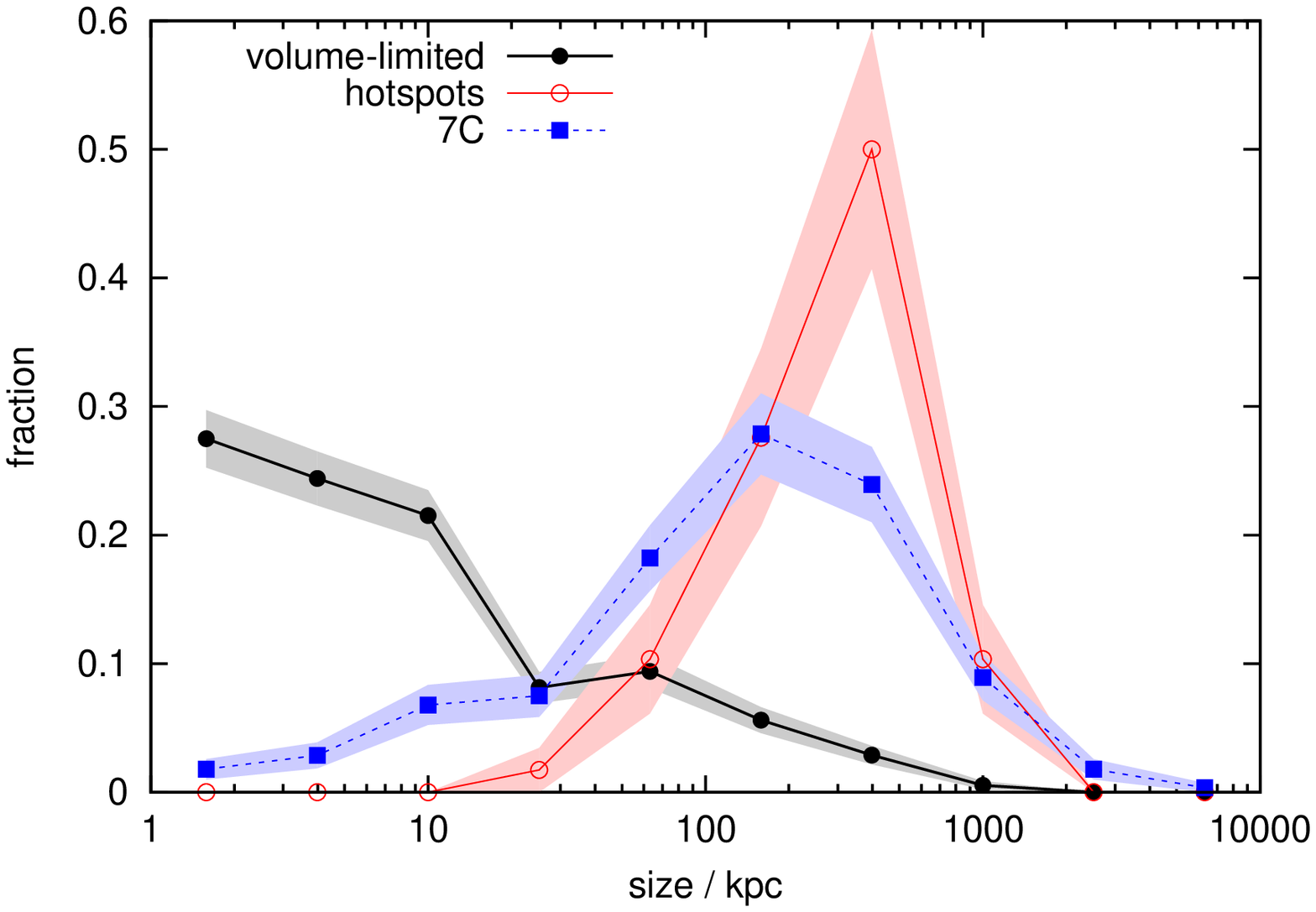}
\end{center}
\caption{Distribution of dynamical jet powers for the volume-limited sample of Shabala et al. 2008 (left panel). The best-fit $Q_{\rm jet} - L_{\rm radio}$ relation from the hotspot sample has been subtracted. The largest sources have dynamical powers broadly consistent with hotspot values (i.e. the black curve peaks near zero). On the other hand, for a given radio luminosity, smaller sources (red and purple triangles, and blue circles) are seen to lie at systematically lower jet powers. The best fit is different for the hotspot and volume-limited samples. This is because the slope of the jet power -- radio luminosity relation depends on the distribution of source sizes in the sample (right panel).}
\label{fig:Qjet_Lradio_sampleSelection}
\end{figure}

We also plot the $Q_{\rm jet} - L_{\rm radio}$ relation for the hotspot sample of Paper~1 (blue dashed line). The two samples (hotspot and volume limited) clearly show very different scalings between radio luminosity and jet power, with slopes of $0.67 \pm 0.05$ and $1.02 \pm 0.02$ respectively. As we show in the top panel of Figure~\ref{fig:Qjet_Lradio_sampleSelection}, the hotspot relationship derived in Paper~1 provides an excellent fit to the largest ($>$200~kpc) sources in the volume-limited sample. This is because the hotspot sample consists mostly of sources of this size (see bottom panel of Figure~\ref{fig:Qjet_Lradio_sampleSelection}). On the other hand, jet powers for the smallest, faintest sources in the volume-limited sample are overpredicted by 0.5 dex.

The difference in the slopes of the $Q_{\rm jet} - L_{\rm radio}$ relations can be understood by considering the effects of the size-luminosity correlation in each sample on Equation~\ref{eqn:LQD_bestFit}. In the hotspot sample, the brightest sources are also the most compact (Figure~\ref{fig:PDtracksHotspots}). On the other hand, in the volume-limited sample it is the largest sources that are brightest (see Figure~12 in Shabala \etal\/ 2008). Thus, for the hotspot sample the exponent of $L_{151}$ in Equation~\ref{eqn:LQD_bestFit} is $<0.8$ while for the volume-limited sample it is $>0.8$. The exact slope depends on the relative numbers of large and small sources, i.e. the size distribution shown in Figure~\ref{fig:Qjet_Lradio_sampleSelection}. Thus, any derived $Q_{\rm jet} - L_{\rm radio}$ relation depends sensitively on sample selection effects. 

We note that in this analysis we excluded all sources smaller than 10~kpc from our volume-limited sample. This is because in dense galaxy cores, source luminosity can actually {\it increase} as the source grows (Alexander 2000). Hence for the smallest ($\lesssim 10$~kpc) sources, low radio luminosity does not necessarily mean low jet power, and our dynamical model will systematically underestimate the jet power. It is worth noting that the number counts of these small sources (Figure~\ref{fig:Qjet_Lradio_sampleSelection}) far exceed those of larger AGN, arguing strongly in favour of a population that never grows to large sizes. These sources are most likely disrupted due to a combination of low jet power and high environment density (Alexander 2000), and will exhibit a higher radio luminosity at a given jet power (see Equation~\ref{eqn:QjetLradio_spIndex}), due to the so-called ``environmental boosting'' (Barthel \& Arnaud 1996). Thus, even with a proper treatment of their evolution, compact sources will have systematically lower $Q_{\rm jet} / L_{\rm radio}$ ratio, resulting in a further steepening of the relation in comparison with samples consisting of only large, powerful radio sources.

\subsection{AGN kinetic power budget}
\label{sec:Qjet_function}

We conclude with a brief discussion of the implications of our findings on the AGN energy budget, and feedback from AGN on galaxies. Adopting a simple scaling relation between radio luminosity and jet power that does not include a measure of source size or age results in significant errors in jet power estimates. For example, Equation~\ref{eqn:LQD_bestFit} shows that calibrating the jet power - radio luminosity relation using the largest ($\gtrsim 500$~kpc) sources results in overpredicting the jet power for the smallest ($10$~kpc) sources by an order of magnitude. Conversely, calibration of the relation using the smallest sources would underpredict jet power for the largest, oldest AGN by a similar amount.

This is important, because understanding the detailed interplay between AGN jets and their surroundings is crucial to quantifying AGN feedback.  Incorrect estimates of AGN jet power will result in mis-modelling of AGN feedback processes. For example, the most powerful radio sources can affect multiple galaxies in a single AGN outburst (Rawlings \& Jarvis 2004; Shabala et al. 2011). Accurate jet power measurements, and interaction of jets with their environment, are crucial to establishing the population of sources which can grow to large enough sizes to affect this feedback. On the other hand, a number of authors (e.g. Best et al. 2006; Shabala \& Alexander 2009; Pope et al. 2012; McNamara \& Nulsen 2012) have argued that frequent low-level feedback from low-luminosity AGN is responsible for preventing excess cooling and star formation in the low-redshift Universe. Accurate jet power measurements in these objects are required to test whether the heating and cooling rates are indeed balanced. Overall, our results indicate that accurate estimates of the integrated kinetic power output of AGN can only be obtained if a measure of radio source ages, such as size or spectral index, is used in addition to their luminosities. 

\section{Conclusions}
\label{sec:conclusions}

We have used new kinetic jet power measurements of Godfrey \& Shabala (2012) to examine the relationship between integrated radio luminosity and jet power in steep-spectrum extragalactic radio sources. We find that the use of radio luminosity as a proxy for jet power under/over-predicts the jet powers for the largest/smallest sources. We provide a new fitting function for jet power as a function of both source size and luminosity,

\begin{eqnarray}
\frac{Q_{\rm jet}}{10^{36} {\rm \,W}} &=& 1.5^{+1.8}_{-0.8}  \left( \frac{L_{\rm 151}}{10^{27}{\rm \,W\,Hz^{-1}}} \right)^{0.8} \nonumber \\
&& \times \left( 1+z \right)^{1.0} \left( \frac{D}{\rm kpc} \right)^{0.58 \pm 0.17}
\end{eqnarray}

Because of this source size effect, any derived radio luminosity -- jet power scaling relation is sensitive to the distribution of source sizes making up the sample. We have illustrated this point by comparing the sample of 3C FR-II radio galaxies used by Godfrey \& Shabala (2012) with the complete volume-limited sample of Shabala et al. (2008). In the hotspot sample, the brightest sources are the most compact; the opposite is true for the volume-limited sample, where the largest FR-II radio sources are most luminous. The different correlations of source size with luminosity in the two samples yield very different slopes for the $Q_{\rm jet} - L_{\rm radio}$ relation, $0.67 \pm 0.05$ and $1.02 \pm 0.02$ respectively. Thus, caution must be taken when using radio source luminosity to infer AGN jet kinetic power.

\appendix

\section{Numerical Constants}

We provide expressions for the normalisation constants discussed throughout the paper. These constants were given in \citet{KA97}, \citet{KDA97} and \cite{KaiserBest07}, but for completeness we recount them here. We define $\Gamma_l$ and $\Gamma_{\rm x}$ to be the adiabatic index of the lobe and external medium respectively, $p$ the energy index of the electron energy distribution $n(\gamma) = k_e \gamma^{-p}$, $R_{\rm T}$ the axial ratio of the source, and the source is assumed to expand into an atmosphere with density profile $\rho(r)=\rho_0 \left( \frac{r}{R_0} \right)^{-\beta}$.

For models used in the paper, we assume values of $R_{\rm T}=2.3$, $p=2.5$, $\beta=1.5$, $\Gamma_{\rm x}=5/3$, $\Gamma_{\rm l}=4/3$, $\gamma_1=1$, $\gamma_2=10^5$, $\nu=151$~MHz.

The constant $c_1$ in Equation \ref{eqn:Ddynamical} is 

\begin{eqnarray}
c_1 &=& \left[  \frac{R_{\rm T}^4}{18 \pi} \frac{(\Gamma_{\rm x} + 1)(\Gamma_l-1)(5-\beta)^3}{9 \left[ \Gamma_l + (\Gamma_l - 1) \frac{R_{\rm T}^2}{2} \right] - 4 - \beta} \right]^{1/(5-\beta)}
\end{eqnarray}
and $c_1=1.08$ for our assumed parameters.

The constant $A_1$ in Equations \ref{eqn:Lradio_u_B} and \ref{eqn:LradioLosslessGeneral} is

\begin{equation} \label{eqn:A_1}
A_1 = \frac{16 \pi^2 r_e}{c} \left( \frac{q}{m_e} \right)^{\frac{(p+1)}{2}} \left( 2 \mu_0 \right)^{\frac{(p+1)}{4}} \frac{C_2(p)}{f(p, \gamma_1, \gamma_2)}, 
\end{equation}
where $r_e$ is the classical electron radius, q and $m_e$ are the electron charge and mass respectively, $\mu_0$ is the permeability of free space, $f(p, \gamma_1, \gamma_2) = \int_{\gamma_1}^{\gamma_2} \gamma^{1-p} d\gamma$, and the function 
\begin{eqnarray}
C_2(p) &=& \frac{3^{p/2}    }{2^{\frac{p+13}{2}} \pi^{\frac{p+2}{2}}} \frac{\Gamma_{\rm fn} \left( \frac{p+1}{4}  \right) \Gamma_{\rm fn} \left( \frac{p}{4} + \frac{19}{12}  \right) \Gamma_{\rm fn} \left( \frac{p}{4} - \frac{1}{12} \right) }{\Gamma_{\rm fn} \left( \frac{p+7}{4}  \right)} \nonumber \\
&=& 2.26 \times 10^{-3} \; \; \mbox{for} \;  p = 2.5 \nonumber
\end{eqnarray} 
where $\Gamma_{\rm fn}(z) = \int_0^\infty t^{z-1} e^{-t} dt$ is the Gamma function \citep[see e.\,g.][]{Worrall09}. For our parameters, $A_1=1.0 \times 10^{-9}$.

The constant $f_p$ in Equation \ref{eqn:u_B} and \ref{eqn:LradioLosslessGeneral}, which we quote directly from Equation A4 of \citet{KaiserBest07}, is

\begin{eqnarray}
f_p &=& \frac{18 c_1^{2(5-\beta)/3}}{(\Gamma_{\rm x} +1)(5 - \beta)^2 R_{\rm T}^2} 
\end{eqnarray} 

The constant $A_2$ in Equation \ref{eqn:LQD_scaling} is

\begin{equation}
A_2 = \left(  \frac{4}{5 + p} \right) {\rm log} \left[   A_1 \pi R_{\rm T}^2 \left( \frac{r f_p}{(r+1)(\Gamma_l - 1)} \right)^{\frac{5+p}{4}}  \nu^{\frac{1-p}{2}} \left(  \rho_0 R_0^\beta \right)^{\frac{5+p}{12}}  \right]
\end{equation} 
where $r=(p+1)/4$. 

For our choice of parameters, $A_2=-5.9$. We note that this expression applies only to the lossless model of radio source evolution, and this value will decrease for realistic evolutionary models.
%$-5.61$

The constant $A_3$ in Equation \ref{eqn:QjetLradio_spIndex} is
\begin{eqnarray}
A_3 &=& \left(  A_1 \pi R_{\rm T}^2 \right)^{-\frac{3}{3-\alpha}}   \left( \frac{r f_p}{(r+1)(\Gamma_l - 1)}  \right)^{-3}   \left(  \rho_0 R_0^\beta \right)^{-\frac{1}{2}}  \nu^{\frac{3 \alpha}{3 - \alpha}}
\label{eqn:A_3}
\end{eqnarray}
where $A_3=8.8 \times 10^{8}$ for our choice of parameters. The value of $A_3$ will increase when losses are taken into account.

\acknowledgements{}
SS thanks the Australian Research Council, and LG thanks Curtin University, for research fellowships. We thank the anonymous referee for a constructive and useful report that has greatly improved the paper. This research has made use of Ned Wright's cosmology calculator \citep{Wright06}; and of the NASA/IPAC Extragalactic Database (NED) which is operated by the Jet Propulsion Laboratory, California Institute of Technology, under contract with the National Aeronautics and Space Administration.

\end{document}